\documentclass[prl,twocolumn,superscriptaddress,showpacs,floatfix]{revtex4}

\usepackage{epsfig,amsmath,amssymb,latexsym}

\usepackage{subfigure}

\begin{document}

\title{Topological insulators with perfect vacancy superstructure
and possible implications for iron chalcogenide superconductors }

\preprint{1}

\author{Xiao-Yong Feng}
 \affiliation{Condensed Matter Group,
  Department of Physics, Hangzhou Normal University, Hangzhou 310036, China}
\author{Hua Chen}
  \affiliation{Zhejiang Institute of Modern Physics and Department of Physics,
  Zhejiang University, Hangzhou 310027, China}
\author{Chao Cao}
 \affiliation{Condensed Matter Group,
  Department of Physics, Hangzhou Normal University, Hangzhou 310036, China}

\author{Jianhui Dai}

\affiliation{Condensed Matter Group,
  Department of Physics, Hangzhou Normal University, Hangzhou 310036, China}

\affiliation{Zhejiang Institute of Modern Physics and Department of
Physics, Zhejiang University, Hangzhou 310027, China}

\email{[Corresponding author:daijh@zju.edu.cn]}

\date{March 11, 2011}

\begin{abstract}
Motivated by the newly-discovered intercalated iron chalcogenide
superconductors, we construct a single orbital tight-binding model
for topological insulators on the square lattice with a perfect
vacancy superstructure. We find that such lattice structure
naturally accommodates a non-vanishing geometry phase associated
with the next-nearest-neighbor spin-orbit interaction. By
calculating the bulk band structures and the finite stripe edge
states, we show that the topological insulator phases can be tuned
at certain electron fillings in a wide range of the model
parameters. The possible implications of these results for the iron
deficient compounds $(A,Tl)_{y}Fe_{2-x}Se_2$ have been discussed.
\end{abstract}


\maketitle

Topological insulators (TIs) are a new type of quantum matter which
have become evident in electronic systems with appropriate lattice
structures and the strong spin-orbit interaction
(SOI)~\cite{QiZhang,HasanKane}. They differ from conventional band
insulators with the non-vanishing ${\cal Z}_2$-odd Kramers states
characterized by the gapless counter-propagating excitations on the
boundary of materials. Due to the topological nature, such gapless
states are robust against weak disorder or interactions. In
particular, when these edge states are in proximity to s-wave
superconductors, the Majorana fermions could be realized due to the
proximity effect \cite{FuKane,Wilczek}, and this promises possible
applications in spintronics and fault-tolerant topological quantum
computation~\cite{Nayak,Kitaev}.

One of the prototype systems for TIs is the Kane-Mele model which
describes tight-binding electrons with the SOI on a two-dimensional
honeycomb lattice such as in graphene ~\cite{Kane1, Kane2}. However,
the SOI turns out to be too small to open a measurable gap in
graphene. Zhang and his collaborators predicted the existence of the
two-dimensional TI in HgTe/CdTe quantum well
structures~\cite{Zhang1} and it was soon confirmed experimentally by
observing the quantized residual conductance caused by edge
states~\cite{Konig}. TIs can be also realized in other lattice
systems such as in the diamond lattice model which generalizes the
2D honeycomb lattice~\cite{Fu1} to three dimensions. As the number
of the lattice structures supporting the TI are limited in
electronic materials, several schemes are proposed for realizing the
topological quantum states in cold-atom systems by taking advantage
of the optical lattice in model engineering~
\cite{Optical1,Optical2,Optical3,Optical4}.

Meanwhile, three-dimensional TIs have been theoretically
predicted~\cite{Theory3D} and experimentally realized in several
realistic materials such as $Bi_{1-x}Sb_x$, $Bi_2Se_3$, $Bi_2Te_3$,
$Sb_2Te_3$~\cite{Exp3D1,Exp3D2,Exp3D3,Exp3D4}, and
$TlBiSe_2$~\cite{Exp3D5}. Recently, the known topological compound
$Bi_2Te_3$ has been shown to exhibit pressure-induced
superconductivity with $T_c\sim 3$ K~\cite{Jin}. The real crystal
structures in the relevant materials can be viewed as, more or less,
the layered honeycomb and triangle lattices, or the distorted ones
like the diamond lattice. It remains a challenge to search for the
topological superconductors, or the topological compounds with
intrinsic bulk superconductivity at the ambient pressure.

\begin{figure}[h]
\includegraphics [width=8.5cm]{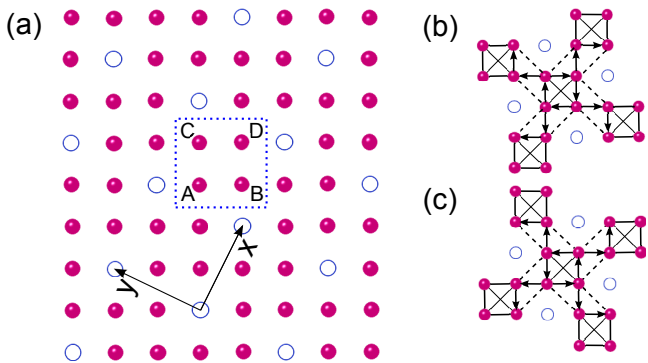}
\caption{(a)The perfect vacancy superstructure (clockwise). The open
dots represent the Fe-vacancies and one of the block is marked by a
dotted square; the x- and y-axes are two directions for the ${\sqrt
5}\times{\sqrt 5}$ bi-partite block lattice. (b) and (c): The
right-handed (clockwise) and left-handed (anti-clockwise) vacancy
orderings, respectively. The solid lines are for hoppings and the
dashed lines are for SOIs. The arrows indicate the permissible paths
for the inter-block SOIs defined by Eq.(2).} \label{lattice}
\end{figure}

Here we suggest that the TIs, or even the topological
superconductors with moderately high transition temperatures at the
ambient pressure, could be possibly realized in materials with the
crystal structures similar to the ones in the 122-type (I4/m)
layered iron chalcogenide superconductors $(A,Tl)_{y}Fe_{2-x}Se_2$
(with $A$ being the alkali atoms)~\cite{Chen,Fang}. One prominent
feature of these newly-discovered materials is that there are iron
vacancies presumably forming the orthorhombic or tetragonal
Fe-vacancy superstructures~\cite{Fang}. Among them, the tetragonal
vacancy superstructure (with the Fe-vacancy density $20\%$) is
perfect and most stable, as within each FeSe layer all Fe-atoms are
three-coordinated, exhibiting the maximal symmetry with either left-
or right-chirality but without breaking the in-plane four-fold
rotational invariance, see in Fig. \ref{lattice}(a). This perfect
superstructure has been clearly observed by a number of
experiments~\cite{JQLi,Bao,Pomjakushin,Ye}.

\begin{figure}[h]
\includegraphics [width=5cm]{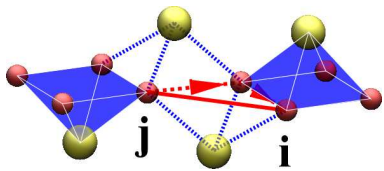}
\caption{An electron at the site ${\bf i}$ hops to the inter-block
n.n.n. site ${\bf j}$ in the presence of a nearby vacancy (with the
right-handed chirality as shown in Fig.\ref{lattice}(b)). This
hopping path takes a non-vanishing geometric phase induced by the
SOI bridged by the Se-atoms (big yellow) located below or above the
Fe-square lattice (small red).}\label{structure}
\end{figure}

Our suggestion is mainly based on the observation that, though being
in the square (or cubic) lattice, the spin-orbit-induced geometric
phase (Berry phase) can be naturally protected by the vacancy
superstructure. Such lattice can be viewed as a bi-partite lattice
with each unit cell containing a minimal square as a
block~\cite{Hua}. In Fig. \ref{structure} the geometric phase for an
electron hopping from the site ${\bf j}$ to the
next-nearest-neighbor (n.n.n.) site ${\bf i}$ with a nearby vacancy
is illustrated. Notice that for the intra-block n.n.n. hopping, the
associated geometric phases coming from the two permissible paths,
both bridged though the Se-atoms in the iron selenides, cancel to
each other as in the conventional square lattice. Hence the chiral
vacancy ordered lattices (Fig.\ref{lattice}(b) or
Fig.\ref{lattice}(c)) preserve the required symmetry as in the
honeycomb or diamond lattices~\cite{Kane1,Kane2,Fu1}.

In the rest of this paper, we will show that the TI phases can
indeed emerge by tuning either band filling or other parameters in a
single-orbital tight-binding model defined on the square lattice
with the perfect vacancy superstructure (Fig.\ref{lattice}(a)). The
tight-binding model Hamiltonian we considered is
\begin{eqnarray}
H_t = \sum_{ij}[t_{ij}c^{\dag}_{i\sigma}c_{j\sigma}+h.c.].
\end{eqnarray}
Where, $c^{\dag}_{i\sigma}$ and $c_{i\sigma}$ are the creature and
annihilation operators for electrons with spin $\sigma$ at the site
${\bf i}$. We also assume $t_{ij}=t_1$ or $t_2$ for the nearest
neighbor (n.n.) or the n.n.n. hoppings within a block, and
$t_{ij}=t'_1$ or $t'_2$ for the corresponding hoppings between the
two n.n. blocks, and no hopping happens among other sites, see in
Fig. \ref{lattice}(b).

Next, we introduce the SOI for the two n.n.n. sites
\begin{eqnarray}
H_\lambda=
i\lambda\sum_{ij}c^{\dag}_{i\sigma}\vec{\sigma}_{\sigma\sigma'}\cdot
(\vec{e}_i\times \vec{e}_j)c_{j\sigma'},
\end{eqnarray}
where $\vec{\sigma}=(\sigma^x,\sigma^y,\sigma^z)$ are the Pauli
matrices, $\vec{e}_j$ and $\vec{e}_i$ are unit vectors along the two
bonds in series when the electron transfers from ${\bf j}$ to ${\bf
i}$.

By taking the block as a unit cell and choosing the coordinate
directions as shown in Fig.\ref{lattice}~\cite{Hua}, we are able to
rewrite the total Hamiltonian $H=H_t+H_\lambda$ in the momentum
space,
\begin{eqnarray}
H=\sum_{\textbf{k}\sigma}F^{\dag}_{\sigma}h_\sigma(\textbf{k})F_{\sigma},
\end{eqnarray}
where
$F^{\dag}_{\sigma}=(c^{\dag}_{A\sigma},c^{\dag}_{B\sigma},c^{\dag}_{C\sigma},c^{\dag}_{D\sigma})$,
with $A$, $B$, $C$, and $D$ being the four sites within a block;
$h_{\uparrow}(\textbf{k})$ is a $4\times4$ traceless Hermitian
matrix with the non-zero matrix elements:
$h_{\uparrow12}=t_1+(t_2'+i\lambda)e^{-ik_x}$,
$h_{\uparrow13}=t_1+(t_2'-i\lambda)e^{-ik_y}$,
$h_{\uparrow14}=t_2+t_1'e^{-ik_y}$,
$h_{\uparrow23}=t_2+t_1'e^{ik_x}$,
$h_{\uparrow24}=t_1+(t_2'+i\lambda)e^{-ik_y}$,
$h_{\uparrow34}=t_1+(t_2'-i\lambda)e^{-ik_x}$; and
$h_{\downarrow}(\textbf{k})=h_{\uparrow}^{T}(-\textbf{k})$.


In the absence of the SOI, the bulk band structures were solved in
Ref.\cite{Hua}, exhibiting the Dirac-like dispersions near some
specific points in the momentum space at the $1/4$ or $3/4$ electron
fillings. Increasing $t_2$ will generally tune the sub-band gaps,
and the mid-band gap may open when the inter-block hopping is
($t'_2$) sufficiently weak. In order to investigate the topological
properties of the system with the SOI, we fix $t_1=t'_1=1$ for
simplicity, and consider $t_2=0, 0.5, 1.0$, corresponding to zero,
moderate, and strong intra-block frustrations, respectively. We find
that for each case either the sub-band gaps or the mid-band gap can
be opened by tuning $\lambda$ at certain electron fillings.

\begin{figure}[h]
\includegraphics [width=6cm]{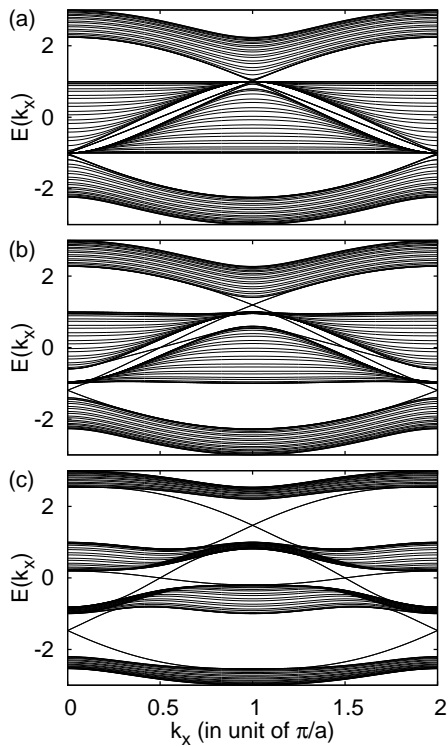}
\caption{(a)$\lambda=0$; (b)$\lambda=0.2$; (c)$\lambda=0.6$. In all
cases, $t_1=1$, $t_1'=1$ and $t_2=t'_2=0$. } \label{nofrustration}
\end{figure}

\begin{figure}[h]
\includegraphics [width=6cm]{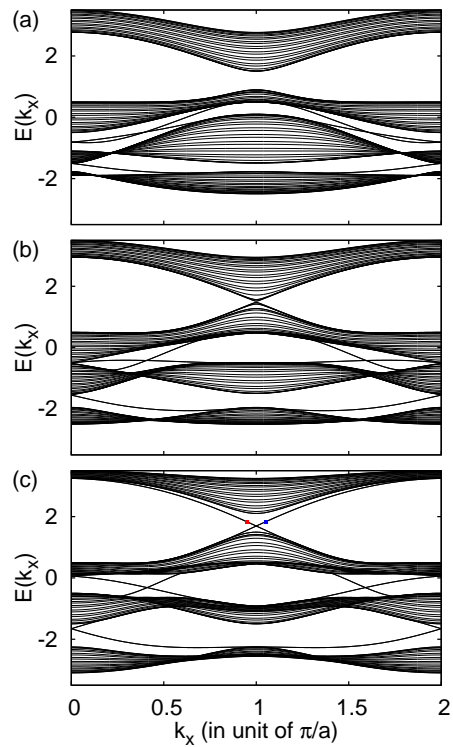}
\caption{(a)$\lambda=0.2$; (b)$\lambda=0.5$; (c)$\lambda=0.8$. In
all cases, $t_1=1$, $t_1'=1$ , $t_2=0.5$ and $t_2'=0$.}
\label{moderate}
\end{figure}

\begin{figure}[h]
\includegraphics [width=6cm]{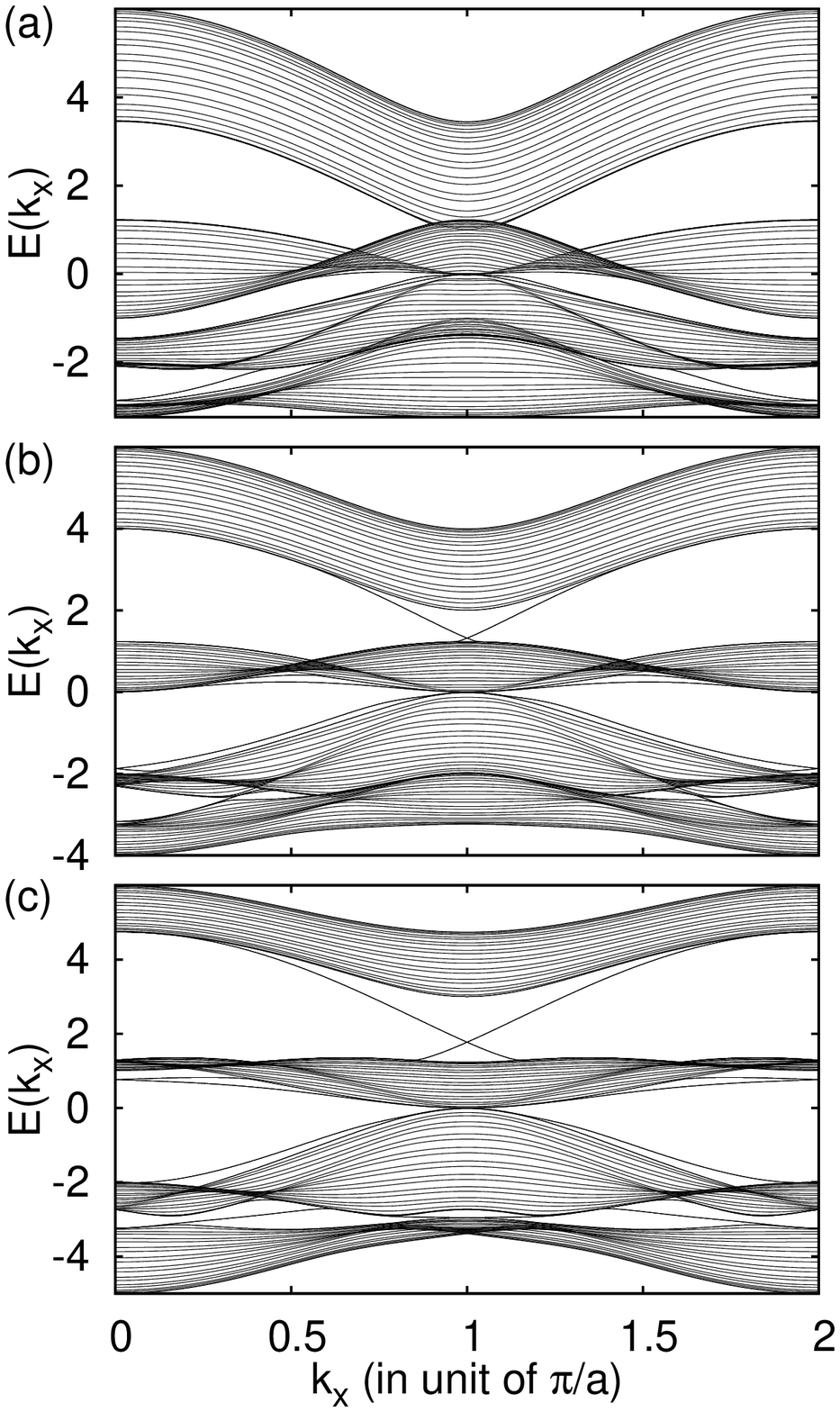}
\caption{(a)$\lambda=0.5$; (b)$\lambda=1$; (c)$\lambda=1.5$. In all
cases, $t_1=t_1'=t_2=t_2'=1$. } \label{strong}
\end{figure}

The topological nature of the insulating phases can be then
investigated by studying the edge states of the strip shaped
lattice. The two edges are cut along the $x$ direction with
sufficient distance (50 unit cells in the our calculations). The
energy spectra are illustrated in
Figs.\ref{nofrustration}-\ref{strong}. For $t_2=0$, the three gaps
among the four bands open gradually with increasing $\lambda$, as
shown in Figs.\ref{nofrustration}(b,c). Meanwhile, various in-gap
states appear within the bulk gaps. Notice that these states are
two-fold degenerate for each momentum $k_x$. We find that a pair of
in-gap states cross within the sub-band gap at the $3/4$-filling at
the time-reversal invariant point $k_x=\pi$. The similar feature is
observed at the $1/4$-filling at another time-reversal invariant
point $k_x=0$. Meanwhile, two pairs of in-gaps states cross within
the mid-band gap at the half-filling near the points $k_x=\pi/2$ and
$3\pi/2$, respectively. As we will show later, all these in-gap
states in Fig.\ref{nofrustration} as well as the in-gap states in
Figs.\ref{moderate},\ref{strong} are true edge states.

Of course, the edge states do not necessarily appear when $t_2$
increases. This is illustrated in the case of $t_2=0.5$ in
Fig.\ref{moderate}(a)($\lambda=0.2$), where no edge state appears at
the $3/4$-filling though the sub-band gap still opens. Notice also
that sometimes the edge states can appear at certain electron
fillings but no corresponding bulk gap opens. When this happens, the
edge states may be coupled to the bulk metallic states upon various
scatterings. We find that the condition for the gapless edge states
to appear within the bulk gap is $\Delta=\lambda+t'_2-t_2>0$, as
explicitly verified in Fig.\ref{moderate} and Fig.\ref{strong} for
moderate and strong frustration cases respectively. In fact, $t'_2$
modifies $\lambda$ and effectively enhances the bulk gap. When
$\lambda=t_2-t'_2$, the Dirac-like dispersion appears at the $3/4$
filling as indicated in Fig.\ref{moderate}(b) where $t'_2=0$,
$t_2=\lambda$($=0.5$).

\begin{figure}[h]
\includegraphics [width=6cm]{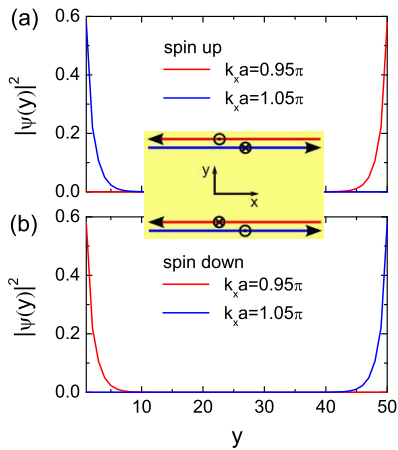}
\caption{The wave function amplitudes of the edge states
corresponding to the red and blue points indicated in Fig.
\ref{moderate}(c). (a)For two spin up states with different momenta;
(b)For two spin down states with different momenta.}
\label{amplitude}
\end{figure}

The properties of the in-gap or edge states can be further clarified
by calculating the amplitudes of the corresponding quasi-particle
wavefunctions. The results for the in-gap states (marked by the red
and blue points in Fig. \ref{moderate}(c)) are shown in Fig.
\ref{amplitude}, where the spin orientation for each state is also
identified. It is shown that each in-gap state is spatially
localized near one of the two edges, while the overlap of the two
wave functions is exponentially small. Focusing on one of the edges,
the two states with opposite spin z-components propagate along the
opposite directions. Hence when the number of such edge state pairs
are odd, they are robust again disorders or weak interactions
preserving the time reversal symmetry\cite{Kane2}. In our model,
such non-trivial TI phases can emerge at the $1/4$ or $3/4$ electron
fillings in a wide parameter region as indicated in Figs.
\ref{nofrustration}(b,c), Fig. \ref{moderate}(c) and Fig.
\ref{strong}(b,c), respectively. The existence of non-trivial TI
phases in our model is compatible with the general classification of
TIs for systems preserving the time reversal symmetry
\cite{Schnyder}.

Finally, we discuss the possible implications of our results for the
newly-discovered iron deficient $(Tl,K)_yFe_{2-x}Se_2$ compounds.
These compounds are essentially multi-orbital materials with sizable
3d-4p hybridizations, and the Fe-3d electron correlation may be
moderate. Recent first principle calculations find that for $x=0.4$
a block-spin N\'eel antiferromagnetic (AFM) ground state can be
stabilized in the perfect vacancy superstructure and a band gap
$\sim 400-500$ meV opens at the Fermi energy when $y=0.8$,
suggesting the Fe$^{2+}$ valence in the charge neutral parent
compound~\cite{CaoDai0.4,LuXiang0.4}. The variation in $y$ just
shifts the chemical potential but does not change the block-spin
ordering, implying that the perfect vacancy structure is robust. The
neutron diffraction experiments suggest that such structure is
stable below 550 K or around~\cite{Bao}. Therefore, one of the most
important ingredients for our model to host TIs, i.e., the perfect
vacancy superstructure which breaks the inversion symmetry by its
chirality, could remain {\it upon various electron or hole dopings}
(by tuning $y$). Another important ingredient, i.e., the SOI, is
intrinsic in various chalcogenide-containing materials as in known
TIs\cite{Exp3D1,Exp3D2,Exp3D3,Exp3D4,Exp3D5}.  The inter-block
frustration hoppings as well as the SOIs in the intercalated iron
chalcogenides are enhanced to due the structure distortion
\cite{CaoDai0.4,LuXiang0.4} (the corresponding Fe-Se-Fe angle is
enlarged), so that the condition $\lambda+t'_2>t_2$ can be
satisfied. It is also suggested that the moderate Coulomb
interaction can induce the SOI through the mechanism of spontaneous
symmetry breaking\cite{Mott}. We can expect that even for relatively
strong Coulomb interaction, the TI phases could be transformed to
the topological AFM/Mott insulator phases~\cite{Balents}. Lastly,
recent experiments have suggested a phase separation~\cite{JQLi} or
co-existence of the block-spin N\'eel AFM order and the
superconductivity~\cite{Bao} in $(Tl,K)_yFe_{2-x}Se_2$. Thus it is
interesting to see whether the TI phases driven by the ordered
vacancies could be in proximity to or co-exist with some kind of
superconducting phases in this or other similar compounds.
Therefore, this class of materials provides a possible new route of
searching for the topological superconductors.

We would like to thank Minghu Fang and Zhuan Xu for helpful
discussions. This work was supported by the NSF of China, the NSF of
Zhejiang Province, the 973 Project of the MOST and the Fundamental
Research Funds for the Central Universities of China (No.
2010QNA3026).


\begin{thebibliography}{99}
\bibitem{QiZhang}X.L. Qi and S.C. Zhang, Physics Today {\bf
63}, 33 (2010).
\bibitem{HasanKane}M.Z. Hasan and C.L. Kane, Rev. Mod. Phys. {\bf
82}, 3045 (2010).
\bibitem{FuKane}L. Fu and C.L. Kane, \prl {\bf 100}, 096407 (2008).
\bibitem{Wilczek}F. Wilczek, {\it Nature Phys.} {\bf 5}, 614 (2009).
\bibitem{Nayak}C. Nayak, S.H. Simon, A. Stern, M. Freedman,
and S. Das Sarma, {\it Rev. Mod. Phys.} {\bf 80}, 1083 (2008).
\bibitem{Kitaev}A.Y. Kitaev, {\it Ann. Phys.} {\bf 303}, 2 (2003).
\bibitem{Kane1}C.L. Kane and E.J. Mele, \prl {\bf 95}, 146802 (2005).
\bibitem{Kane2}C.L. Kane and E.J. Mele, \prl {\bf 95}, 226801 (2005).
\bibitem{Zhang1}B.A. Bernevig, T.L. Hughes and S.C. Zhang, Science \textbf{314},
1757(2006).
\bibitem{Konig}M. K$\ddot{o}$nig, {\it et al.}, Science \textbf{318}, 766 (2007).
\bibitem{Fu1}L. Fu, C. L. Kane, and E. J. Mele, \prl{\bf 98}, 106803
(2007).
\bibitem{Theory3D}H. Zhang, {\it et al.}, Nature Phys. \textbf{5}, 438
(2009).
\bibitem{Exp3D1}D. Hsieh {\it et al.}, Nature \textbf{452}, 970 (2008).
\bibitem{Exp3D2}Y. Xia, {\it et al.}, Nature \textbf{5}, 398 (2008).
\bibitem{Exp3D3}Y.L. Chen {\it et al.}, Science \textbf{325}, 178 (2008).
\bibitem{Exp3D4}D. Hsieh {\it et al.}, \prl{\bf 103}, 146401 (2009).
\bibitem{Exp3D5}K. Kuroda et al, \prl{\bf 105}, 146801 (2010).
\bibitem{Optical1}C. Wu, \prl {\bf 101}, 186807 (2008).
\bibitem{Optical2}L. B. Shao {\it et al.}, \prl{\bf 101}, 246810 (2008).
\bibitem{Optical3}T. D. Stanescu, {\it et al.}, \pra {\bf 79}, 053639
(2009).
\bibitem{Optical4}M. Kargarian and G.A. Fiete, \prb{\bf 82}, 085106 (2010).
\bibitem{Jin}J.L. Zhang {\it et al.}, PNAS {\bf 108}, 24 (2011).
\bibitem{Chen}J. Guo, {\it et al.}, \prb {\bf 82}, 180520(R) (2010).
\bibitem{Fang}M. Fang, {\it et al.}, arXiv:1012.5236v1 (2010).
\bibitem{JQLi}Z. Wang {\it et al.}, arXiv:1101.2059 (2011).
\bibitem{Bao}W. Bao {\it et al.}, arXiv:1102.0830 (2011).
\bibitem{Pomjakushin}V.Y. Pomjakushin {\it et al.}, arXiv:1102.3380
(2011).
\bibitem{Ye}F. Ye {\it et al.}, arXiv:1102.2882 (2011).
\bibitem{Hua}H. Chen, C. Cao, and J. Dai, arXiv:1102.4074 (2011).
\bibitem{Schnyder}A.P. Schnyder {\it et al.}, AIP Conf. Proc. {\bf
1134}, 10 (2009).
\bibitem{CaoDai0.4}C. Cao and J. Dai, arXiv:1102.1344 (2011).
\bibitem{LuXiang0.4}X.W. Yan, M. Gao, Z.Y. Lu, and T. Xiang,
arXiv:1102.2215 (2011).
\bibitem{Mott}S. Raghu, X.L. Qi, C. Honerkamp, S.C.
Zhang, \prl {\bf 100}, 156401 (2008).
\bibitem{Balents}D.A. Pesin and L. Balents, Nature Phys. {\bf 6}, 376 (2010).
\end{thebibliography}
\end{document}